\documentstyle[11pt,epsfig,amssymb]{article}

\begin{document}
\setlength{\parskip}{0.45cm}
\setlength{\baselineskip}{0.75cm}

\vspace{1cm}
%\begin{flushright}
%CERN-TH/99-01 \\
%{\tt astro-ph/9901004}\\
%\end{flushright}
%
\begin{center}
{\bf
Galactic Origin of the Alleged Extragalactic Gamma-Ray Background 
Radiation}\\
\vspace{0.3cm}
\large Arnon Dar\\
\normalsize  Technion, Israel Institute of Technology,
Haifa 32000, Israel\\ 
% %%%%%%%%%%%%%%%%% 
{\bf Abstract} \\ 
\end{center}
\vspace*{-0.5cm} 
\noindent 
The diffuse gamma-ray background radiation (GBR) at high Galactic
latitudes could be dominated by inverse Compton scattering (ICS) of cosmic
ray (CR) electrons on the cosmic microwave background radiation (CBR) and
on starlight (SL)  in an extended CR halo of our galaxy. Assuming the
locally observed CR electron spectrum beyond a few GeV, which follows from
equilibrium between Fermi acceleration and radiative cooling, to be a
universal spectrum throughout our galaxy and its CR halo, we reproduce the
observed spectral index, the intensity and the angular dependence of the
GBR, in directions away from the galactic disk and centre, without
recourse to hypothetical extragalactic sources.

{\bf 1. The GBR:}
The existence of an isotropic, diffuse extragalactic gamma background
radiation (GBR) was first suggested by data from the SAS 2 satellite
(Thompson \& Fichtel 1982) and claimed to be confirmed by the EGRET
instrument on the Compton Gamma Ray Observatory (e.g. Sreekumar et
al.~1998).  The alleged extragalactic diffuse GBR is the diffuse emission
observed by EGRET by masking the galactic plane at latitudes $\rm{|b|\le
10^o}$, as well as the galactic centre at $\rm{|b|\le 30^o}$ for
longitudes $\rm{|l|\le 40^o}$, and by extrapolating to zero column
density, to eliminate the $\pi^0$ and bremsstrahlung contributions to the
observed radiation and to tame the model-dependence of the results.
Outside the mask, the GBR flux integrated over all directions in the
observed energy range of ${\rm 30}$ MeV to ${\rm 120~GeV}$, shown in
Fig.~1, is well described by a power law: 
\begin{equation}
{\rm {dF_\gamma\over dE}\simeq (2.74\pm 0.11)\times 10^{-3} \left
[E\over MeV \right]^{-2.10\pm 0.03}
~cm^{-2}~s^{-1}~sr^{-1}~MeV^{-1}}\; .
\label{photons}
\end{equation}
The overall magnitude
in Eq.~(\ref{photons}) is sensitive to the model used to subtract
the foreground (Sreekumar et al.~1998; Strong et al.~1998), but the  
spectral index is not.
The EGRET data are given in Sreekumar et al.~(1998) for 36 $\rm{(b,l)}$ domains, 9 values for each half-hemisphere.
The spectral index is, within errors, extremely directionally uniform,
as shown in Fig.~2, where we have plotted the EGRET results
as functions of $\theta$, the observation angle
relative to the direction to the galactic centre 
(${\rm \cos\theta=\cos\,[b]\, \cos\,[l]}$).
The normalization 
is less homogeneous, but in directions well above the galactic disk and 
away from the galactic-centre region it
has been found to be consistent with
a normal distribution around the mean value:
thus the claim of a possible extragalactic origin (Sreekumar et al. 1998).

The origin of this diffuse GBR is still unknown.
The published candidate sources range from the
quite conventional to the decisively speculative.
Perhaps the most conservative hypothesis is that the GBR
is extragalactic, and originates from active galaxies (Bignami et al.~1979;
Kazanas \& Protheroe 1983; Stecker \& Salamon 1996).
The fact that blazars have a $\gamma$-ray spectrum with an average index
$2.15\pm 0.04$, compatible with that of the GBR, supports
this hypothesis (Chiang \& Mukerjee 1998). 
The possibility has also been discussed
that Geminga-type pulsars, expelled into
the galactic halo by asymmetric supernova explosions,
be abundant enough to explain the GBR (Dixon et al. 1998;
Hartmann 1995). More exotic hypotheses include a baryon-symmetric
universe (Stecker et al.~1971), now excluded (Cohen et al.~1998), primordial black hole evaporation (Page \& Hawking 1976; Hawking 1977),  
supermassive black holes
formed at very high redshift (Gnedin \&  Ostriker 1992), annihilation of
weakly interactive  big-bang remnants (Silk \& Srednicki 1984; Rudaz  
\& Stecker 1991), and a long etc.
 
The EGRET GBR data in directions above the galactic disk and centre show a
significant deviation from isotropy, correlated with the structure of our
galaxy and our position relative to its centre (Dar et al.~1999).  In
Fig.~4 we have plotted, as a function of $\theta$, the EGRET GBR
counting-rate above 100 MeV. This figure clearly shows, in three out of
the four quarters of the celestial sphere, an increase of the counting
rate towards the galactic centre. The ``reduced'' $\chi^2$ per degree of
freedom for a constant flux is 2.6: very unsatisfactory. A best fit of the
form ${\rm F=F_0+F_1\,(1-\cos\theta)}$ yields ${\bar\chi^2=1.3}$, a very
large amelioration (for higher polynomials in $\cos\theta$ the
higher-order coefficients are compatible with zero: the fit does not
significantly improve). Note also that at angles with $\cos\theta$ larger
than its mean value $\langle\cos\theta\rangle\! =\! 0.0246$ (${\rm
\theta<88.6^o}$), 10 out of the 12 data points are above the average flux,
while at angles with ${\rm \theta>88.6^o}$, 18 out of the 24 data points
are below the average. The probability for a uniform distribution to
produce this large or a larger fluctuation is $1.5\times 10^{-4}$. Even in
directions pointing to the galactic disk and the galactic centre, EGRET
data on $\gamma$-rays above 1 GeV show an excess over the expectation from
galactic cosmic-ray production of $\pi^0$'s (Pohl \& Esposito 1998).
Electron bremsstrahlung in gas is not the source of the 1--30 MeV
inner-Galaxy $\gamma$-rays observed by COMPTEL (Strong et al.~1997), since
their galactic latitude distribution is broader than that of the gas.
These findings also imply that inverse Compton scattering may be much more
important than previously believed (Strong \& Moskalenko 1998; Moskalenko
and Strong, 2000; Dar et al.~1999).

Thus, it appears that the EGRET data advocates a local (as opposed to
cosmological) origin for the GBR. Indications of a large galactic
contribution to the GBR at large latitudes were independently found by
Dixon et al. (1998) by means of a wavelet-based ``non-parametric''
approach that makes no reference to a particular model.  Strong \&
Moskalenko (1998) and Moskalenko \& Strong (2000) also found that the
contribution of inverse Compton scattering of galactic cosmic ray
electrons to the diffuse $\gamma$-ray background is presumably much larger
than previously thought. 

{\bf 2. Galactic Origin of the GBR:}
In recent publications we went one step further (Dar et al.~1999; Dar \& De 
R\'jula 2000) and
explored in detail the possibility  that the diffuse
gamma-ray background radiation at high galactic latitudes could be
dominated by inverse Compton scattering of cosmic ray (CR) electrons on
the cosmic microwave background radiation and on starlight from our own
galaxy.  We made three very simple assumptions:  (1) The locally-measured
electron spectrum, shown in Fig.~4, which
is well fitted, from  ${\rm E\sim 10~GeV}$  to $\sim 2$ TeV  by,
\begin{equation}
{\rm {dF_e\over dE}\simeq (2.5\pm 0.5)\times 10^5 \left
[E \over MeV \right]^{-3.2\pm 0.10}
~cm^{-2}~s^{-1}~sr^{-1}~MeV^{-1}},
\label{electrons}
\end{equation}
is representative of its average form throughout the
Galaxy and its extended CR halo. (2) Above a few GeV the acceleration of
CR electrons is in equilibrium with their cooling by inverse Compton 
scattering (ICS) on starlight and on the microwave background radiation. 
(3) The GBR is produced by ICS of CR electrons on CBR and starlight
photons in our Galaxy and in external galaxies. 

Assumptions (1) and (2) allow us to derive directly, the GBR spectral
index from the observed electron index:  $\rm
\beta_\gamma=(\beta_e+1)/2=2.10\pm 0.05$. It agrees with the GBR index as
observed by EGRET, $\rm \beta_\gamma=2.10\pm 0.03$. This index is universal,
independent of whether the ICS took place in our Galaxy or in external 
galaxies.

The two dominant contributions to the GBR within our model are inverse
Compton scattering of Galactic CR-electrons off the cosmic background
radiation and starlight. There is a small additive effect of ICS from 
sunlight. The contribution from external galaxies, is also sub-dominant. 
The ICS spectrum --a cumbersome convolution (Felten \& Morrison 1966) of
a CR power spectrum $\rm dn_e/dE=A\, E^{-{\beta_e}}$
with a photon thermal distribution-- can be 
approximated very simply (Dar \& De R\'ujula 2000) by
\begin{equation}
{\rm {dF^i\gamma\over dE}=
{N_i(b,l)~\sigma_{_T}\,A\over 2}~
\left[{4\,\epsilon_i\,MeV\over 3\,m_e^2c^4}\right]^{{\beta_e-1\over 2}}
\,\left[{E\over MeV} \right]^{-{\beta_e+1\over 2}}}\, ,
\label{ICSphotpred}
\end{equation}
where $\rm \sigma_T=0.65\times 10^{-24}\, cm^{-2}$ is the Thomson 
cross section, $\rm \epsilon_i $  are the mean photon energies of starlight
and of the CBR and
${\rm N_i(b,l)}$ are the effective CR electron column density 
resulting from the convolution of the space distribution of CR electrons
with those of starlight and of the CBR in the direction (b,l). 
Simple analytical expressions for them are given in Dar \& De R\'ujula 
(2000).

Our predictions of the intensity and
the angular dependence of the GBR are shown in Figs. 5, 6. 
Like that for the spectral index, they are in excellent agreement with
those observed for the alleged extragalactic GBR, leading to the conclusion 
that the GBR could be dominated by the emission from our own galaxy. 

The only non-conventional aspect of our model is that, in order to
reproduce the observed intensity of the GBR, we must assume the scale
height of our galaxy's CR-electron distribution to be almost twice the
traditionally-accepted upper limit.  Such a large scale height is not in
contradiction with radio synchrotron-emission from our galaxy if the
galactic disk and its magnetic field are embedded in a larger magnetic
halo with a much weaker field. 

{\bf 3. Predictions:} The main predictions specific to our scenario are:\\
(1) The GBR should reflect the
asymmetry of our off-centre position in the Galaxy.\\
(2) Very nearby star-burst Galaxies, such as M82, 
and radio galaxies with presumably large CR production rates, such as 
Cygnus A, may be visible in gamma rays.\\ 
(3) The GBR  spectrum should not have the sharp cutoff,
above ${\rm E\sim 100}$ GeV,
expected (Salamon \& Stecker 1998) for cosmological sources.
But it should nonetheless steepen around 10--100 GeV, because of
the anticipated ``knee'' in the electron spectrum and of the
energy-dependence of the Klein-Nishina cross section.\\

\noindent
These features of our scenario should  be testable
when the next generation of cosmic-ray and
$\gamma$-ray satellites (AMS-02 and GLAST) are
operational, hopefully by 2005.

ACKNOWLEDGEMENTS\\
This talk is based on a research done in collaboration with A. De R\'ujula.
It was supported in part
by the Helen Asher Fund for Space Reseach and by the
V.P.R. Fund - Steiner Research Fund at the Technion.

\noindent
REFERENCES\\
Barwick S. W. et al., 1998, ApJ, 498, 779 \\
Bignami G. et al., 1979, ApJ, 232, 649\\
Cohen A., De R\'ujula A., Glashow S. L., 1998, ApJ, 495, 539\\
Dar A., De R\'ujula A., Antoniou N., 1999, astro-ph/9901004, in press\\
Dar A., De R\'ujula A., astro-ph/0005080, MNRAS in press\\
Dixon D. D. et al., 1998, New Astron. 3, 539\\
Evenson P., Meyers P., 1984,  J. Geophys. Res., 89 A5, 2647 \\
Felten J. E., Morrison P., 1966, ApJ, 146, 686\\
Ferrando P. et al., 1996, A\&A 316, 528 \\
Golden R. L. et al., 1984, ApJ, 287, 622\\
Golden R. L., et al., 1994, ApJ, 436, 739 \\
Hartmann D. H., 1995, ApJ, 447, 646\\
Kazanas D., Protheroe J. P., 1983, Nature, 302, 228\\
Moskalenko I. V., Strong A. W., Reimer O., 1998, astro-ph/9811221\\
Moskalenko I. V., Strong A. W., 2000, ApJ, 528, 357\\
Nishimura  J. et al., 1980, ApJ, 238, 394 \\
Page D. N., Hawking S. W., 1976, ApJ, 206, 1 \\
Pohl M., Esposito J. A., 1998, ApJ, 507, 327\\
Prince T. A., 1979,  ApJ, 227, 676\\
Rudaz  S.,  Stecker  F. W., 1991, ApJ, 368, 40\\
Salamon  M. H.,  Stecker F. W., 1998, ApJ, 493, 547\\
Silk J., Srednicki M., 1984,  PRL, 53, 264\\
Sreekumar P. et al., 1998, ApJ, 494, 523\\
Stecker F. W., Salamon M. H., 1996, ApJ, 464, 600\\
Stecker F. W., Morgan D. L., Bredekamp J., 1971, PRL, 27, 1469\\
Strong A., Moskalenko I. V., 1998, ApJ, 509, 212\\
Strong  A. W. et al., 1997, in {\it Proceedings of the 4th Compton 
Symposium},  AIP, 410, 1198\\
Strong A. W., Moskalenko I. V., Reimer O., 1998, astro-ph/9811296\\
Tang K. K., 1984,  ApJ, 278, 881 \\
Thompson  D. J., Fichtel C. E., 1982, A\&A, 109, 352\\

%\end{thebibliography}

\newpage

%%%%%%%%%%%%%%%%%%%%%%%%%%%%%%2
\newpage
\begin{figure}
%\plotone{gbrfig1.eps}
\begin{center}
\vspace*{-1.6cm}
\hspace*{-1cm}
\epsfig{file=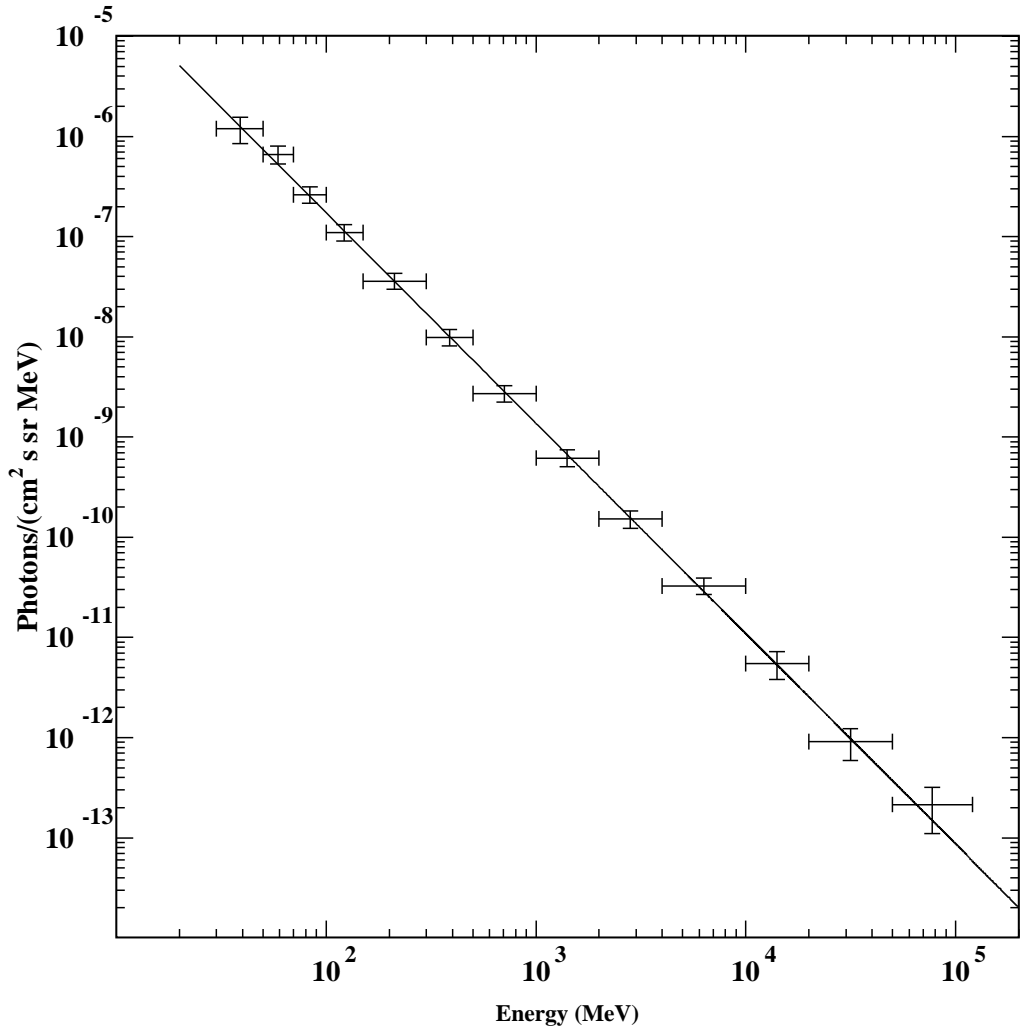,width=15cm}
\caption{Comparison between the spectrum  of
the GBR, measured by EGRET (Sreekumar et al.~1998),
and the prediction for ICS of starlight and the CMB by CR
electrons. The slope is our central prediction, the normalization
is the one obtained for ${\rm h_e}= 20$ kpc, ${\rm \rho_e}= 35$ kpc.}
\vspace*{-0.5cm}
\end{center}
\end{figure}

%%%%%%%%%%%%%%%%%%%%%%%%%3
%____________________________________________________________
\begin{figure}[t]
\begin{tabular}{cc}
\hskip -0.5truecm
\epsfig{file=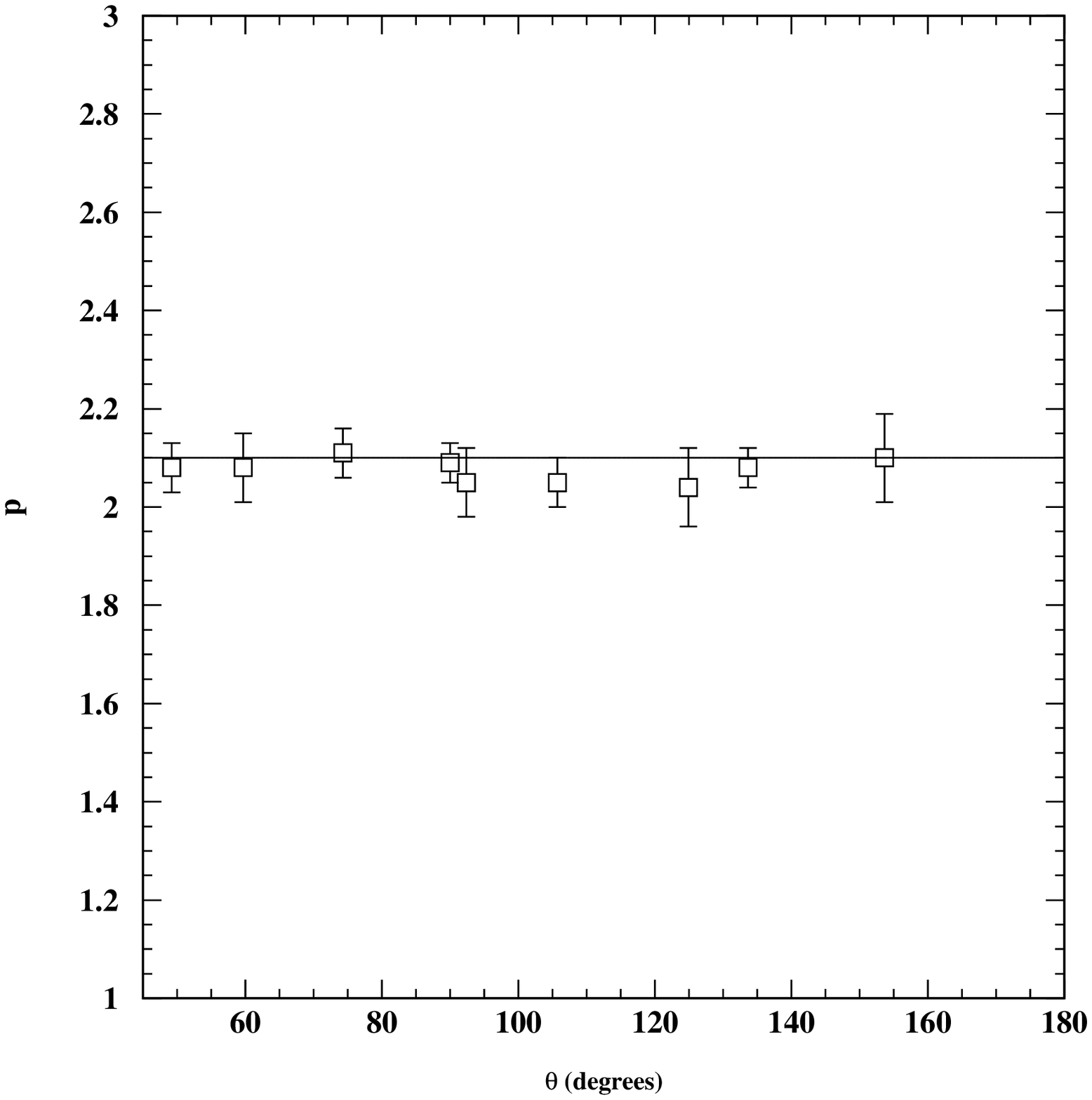,width=8cm} &
\hskip -0.5truecm
\epsfig{file=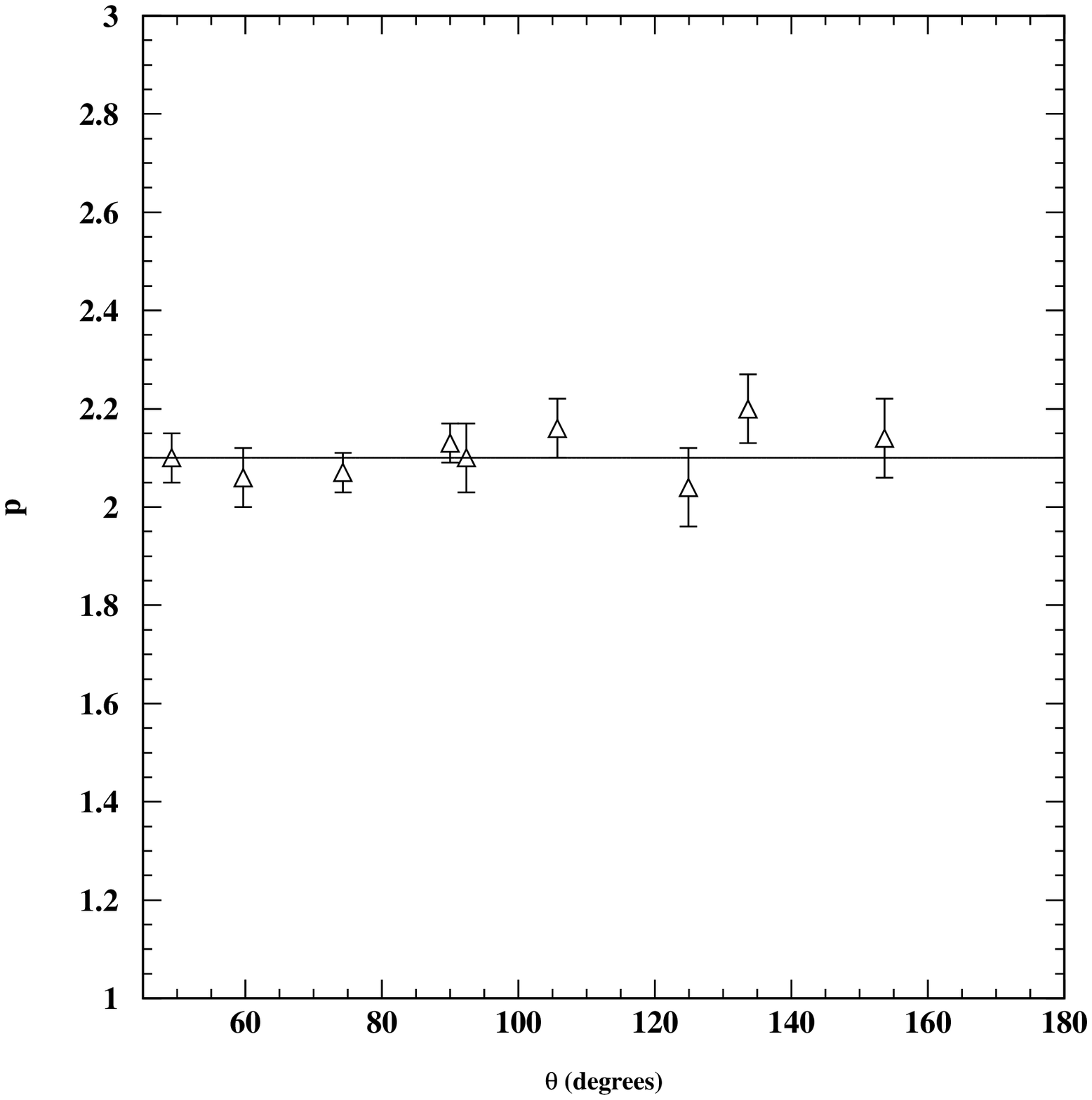,width=8cm}
\vspace{-0.5cm}\\
%\hskip 0.2truecm
{\small (a)}            &
\hskip -0.2truecm
{\small (b)} \\
\epsfig{file=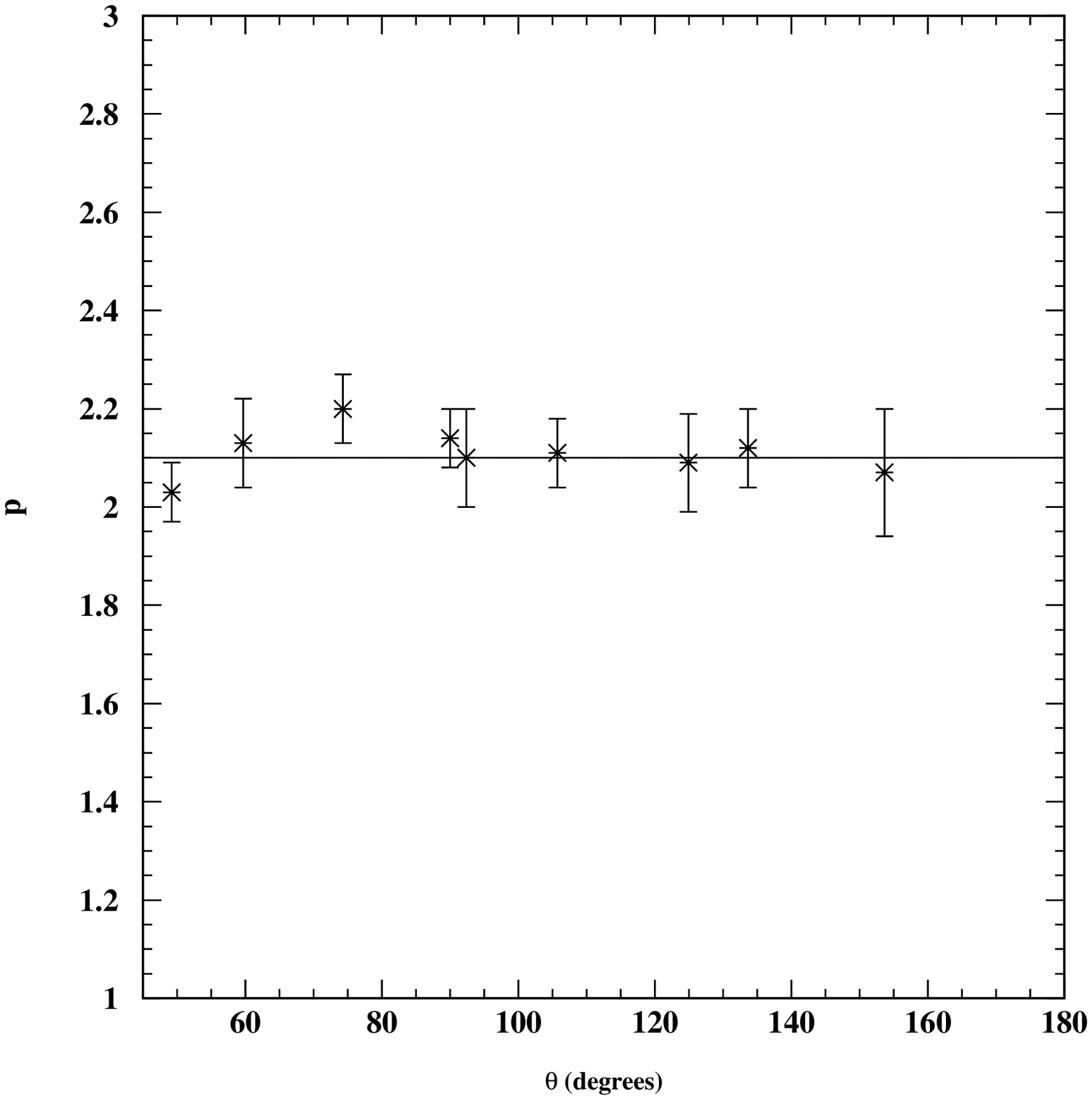,width=8cm} &
\hskip -0.5truecm
\epsfig{file=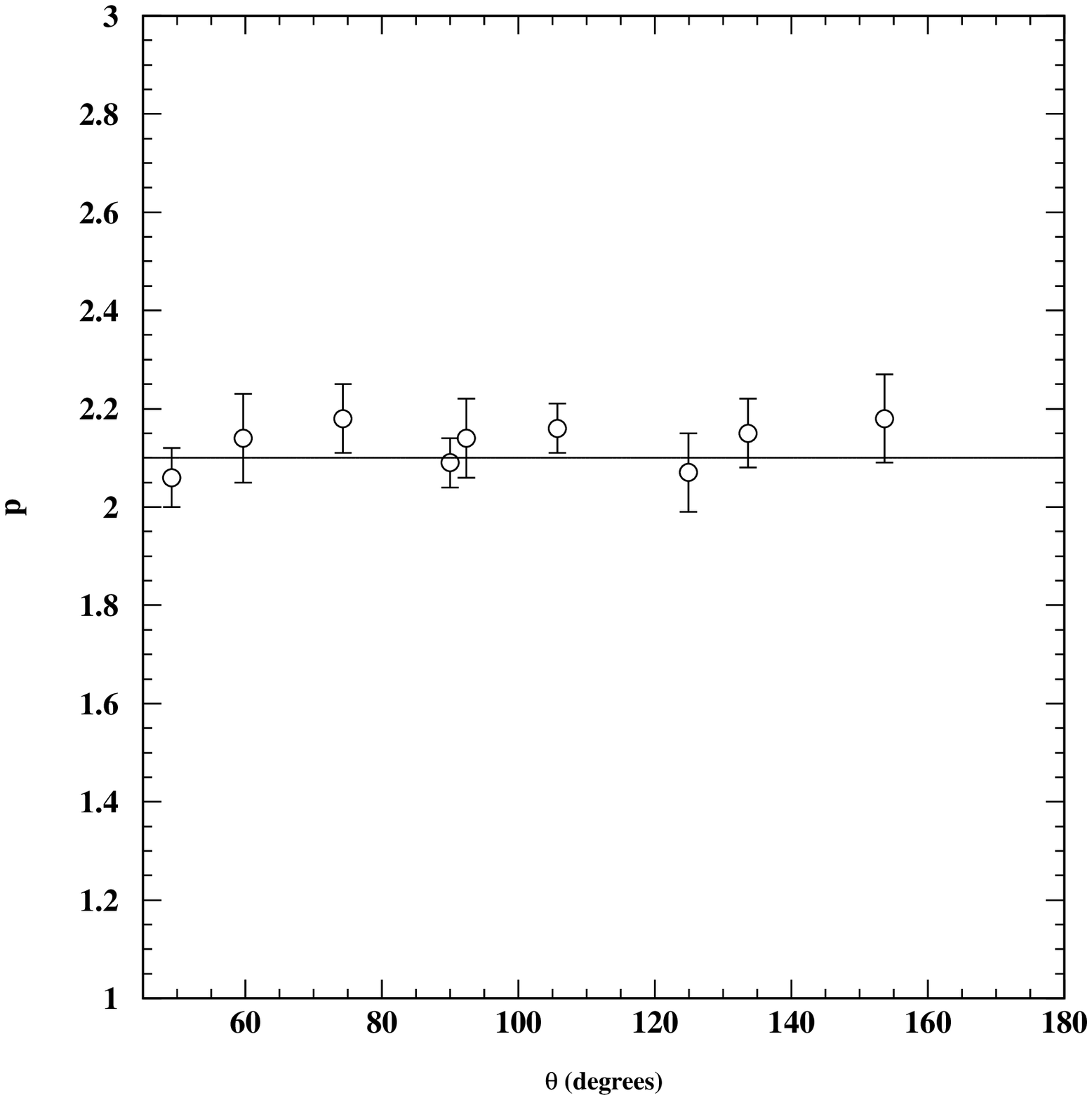,width=8cm}
\vspace{-0.5cm}\\
\hskip 0.2truecm
{\small (c)}            &
\hskip -0.20truecm
{\small (d)}
\end{tabular}
\caption{{
EGRET data on the GBR spectral
index  as a function of  $\theta$
the angle away from the direction of the galactic center.
The line is the predicted spectral index.
The various plots correspond to the individual half-hemispheres.
(a)
${\rm b> 0}$, ${\rm l> 0}$.
(b)
${\rm b> 0}$, ${\rm l< 0}$.
(c)
${\rm b< 0}$, ${\rm l> 0}$.
(d)
${\rm b< 0}$, ${\rm l< 0}$.}}
\label{fig:indexes}
\end{figure}
%__________________________________________________________

%____________________________________________________________
\begin{figure}[t]
\begin{tabular}{cc}
\hskip -0.5truecm
\epsfig{file=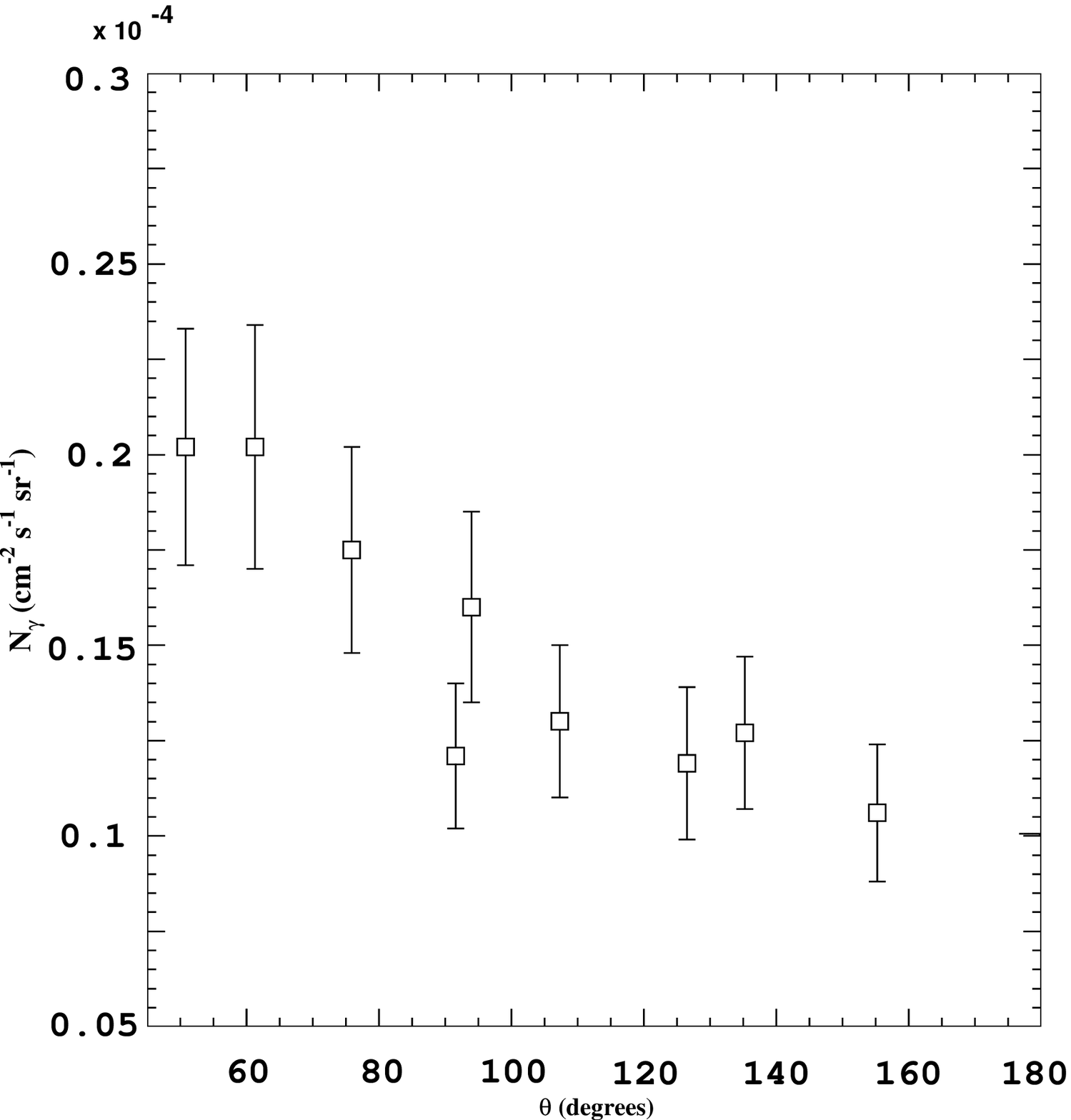,width=7cm} &
\hskip 0.5truecm
\epsfig{file=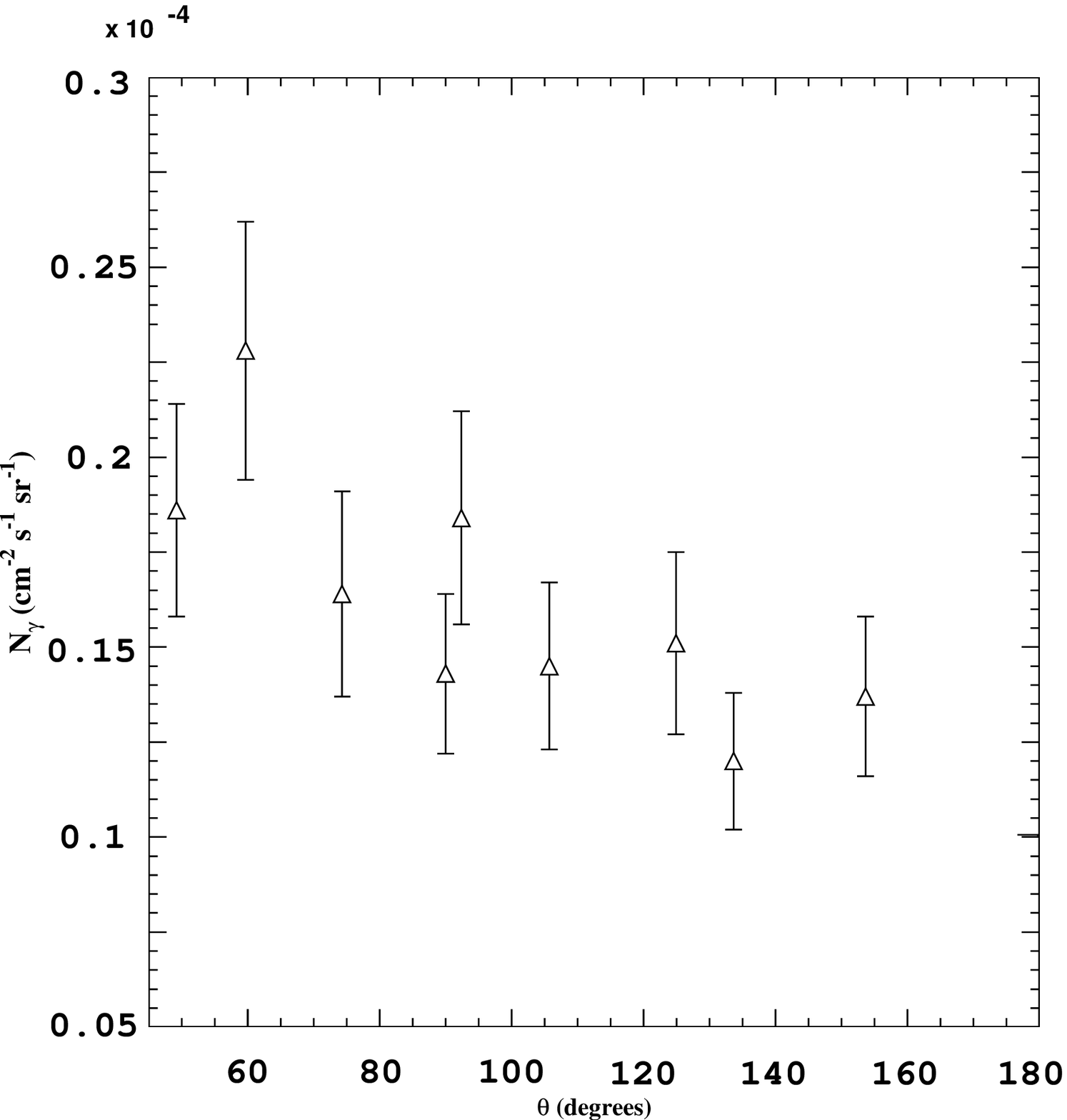,width=7cm}
\vspace{-0.5cm}\\
{\small (a)}            &
\hskip 1.2truecm
{\small (b)} \\
\vspace{-0.9cm}\\
\hskip -0.5truecm
\epsfig{file=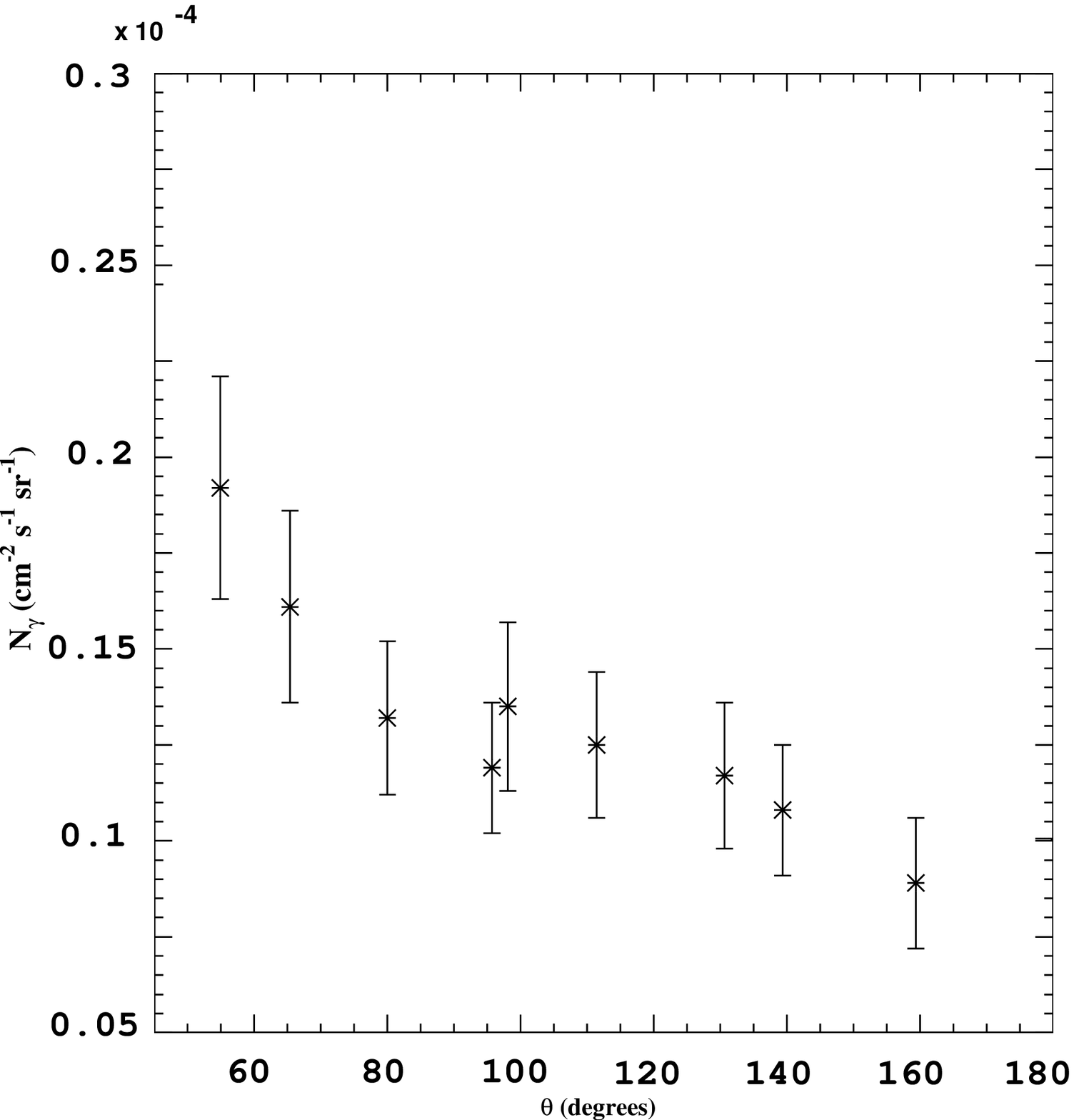,width=7cm} &
\hskip 0.5truecm
\epsfig{file=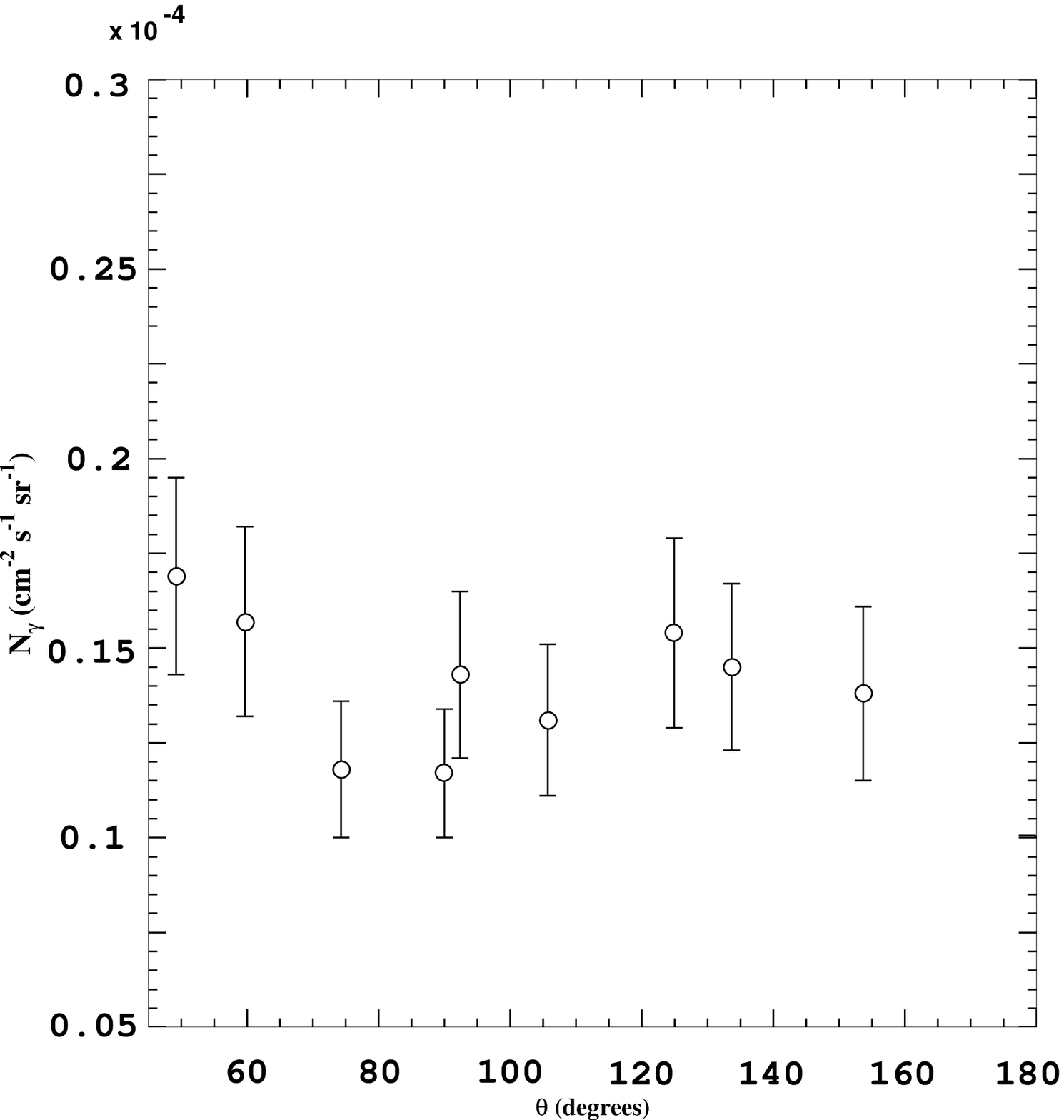,width=7cm}
\vspace{-0.5cm}\\
\hskip 0.2truecm
{\small (c)}            &
\hskip 1.2truecm
{\small (d)}
\end{tabular}
\caption{{EGRET data, organized as in Fig.~2,
for the dependence on $\theta$ of the GBR intensity 
above 100 MeV. }}
\label{fig:intensities1}
\end{figure}
%__________________________________________________________ 

%%%%%%%%%%%%%%%%%%%%%%%%%%%%1
\begin{figure}
\begin{center}
\vspace*{-1.6cm}
\hspace*{-1cm}
\epsfig{file=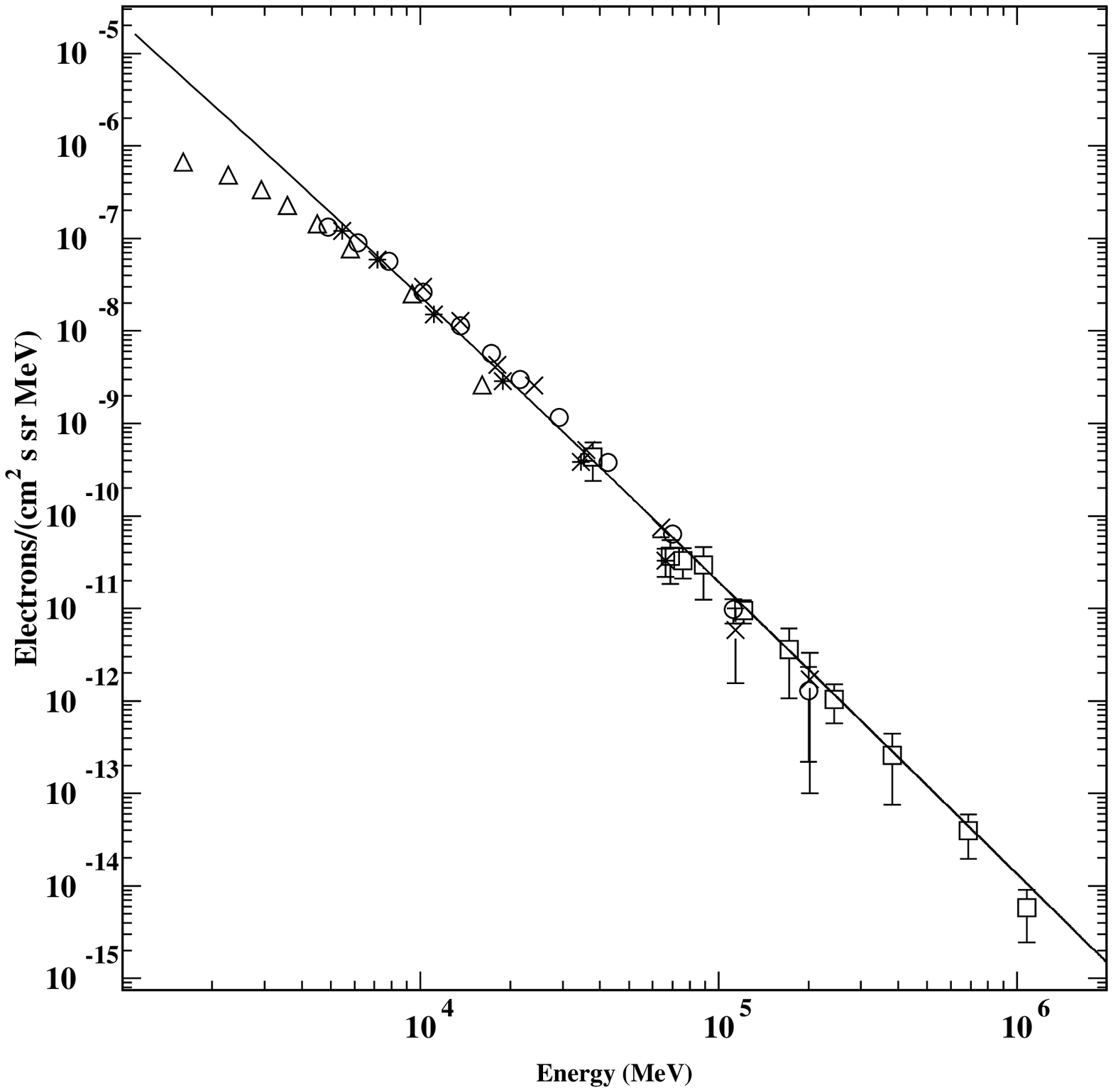,width=15cm}
%\vspace*{-14.6cm}
\caption{The primary cosmic-ray electron spectrum
(Evenson \& Meyers 1984; Golden et al.~1994; Ferrando et al.~1996) as
measured by Prince 1979 [crosses]; Nishimura et al.~1980 [squares];
Tang 1984 [circles]; Golden et al.~1984 [triangles];  Barwick et
al.~1998 [stars]. The slope is the prediction, the
magnitude is normalized to the data.}
\vspace*{-0.5cm}
\end{center}
\end{figure}

%________________________________________________________

\begin{figure}
\begin{center}
\vspace*{-1.6cm}
\hspace*{-1cm}
\epsfig{file=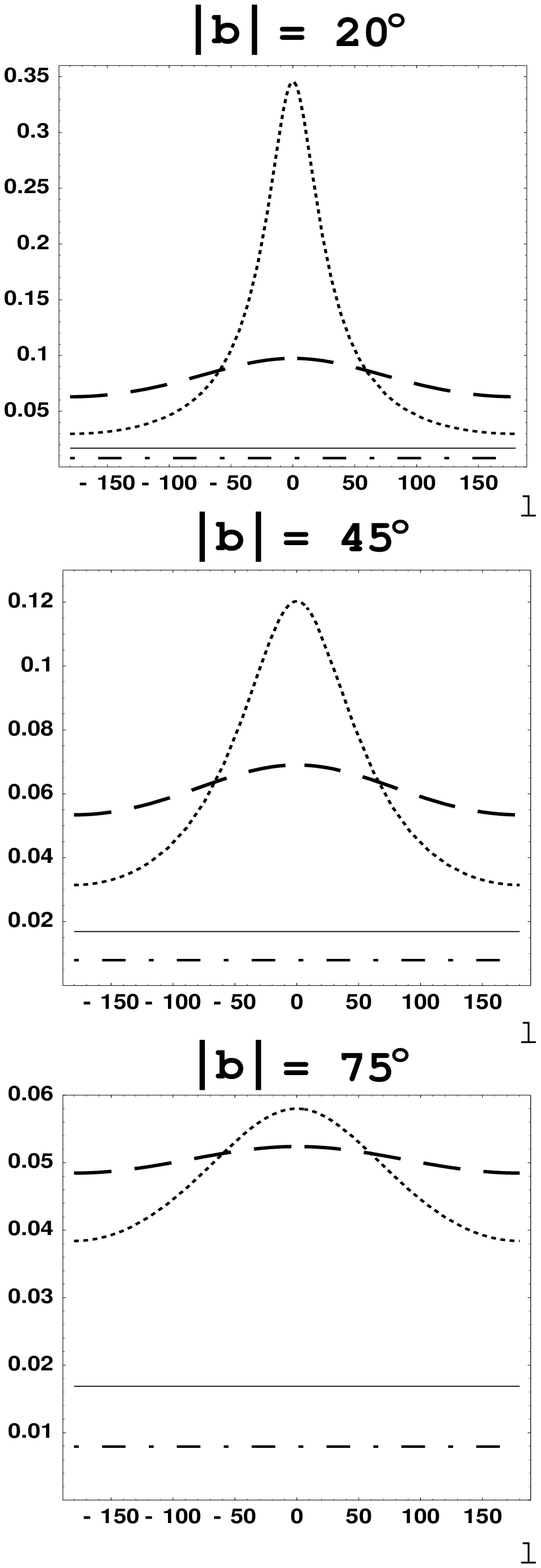,width=6cm}
%\vspace*{-14.6cm}
\caption{Contributions to the GBR flux above 100 MeV
as functions of longitude l, at fixed latitude b, from ICS
of starlight (dotted), and CMB in our galaxy (dashed); from
the total ICS from external galaxies (continuous), and
from sunlight (dot-dashed). The
vertical scale is
$10^4$ times the number of photons/${[\rm cm^2\,s\,sr]}$.
 The results are
for ${\rm h_e=20}$ kpc, ${\rm \rho_e=35}$ kpc.}
\vspace*{-0.5cm}
\end{center}
\end{figure}

\begin{figure}
\begin{center}
\vspace*{-1.6cm}
\hspace*{-1cm}
\epsfig{file=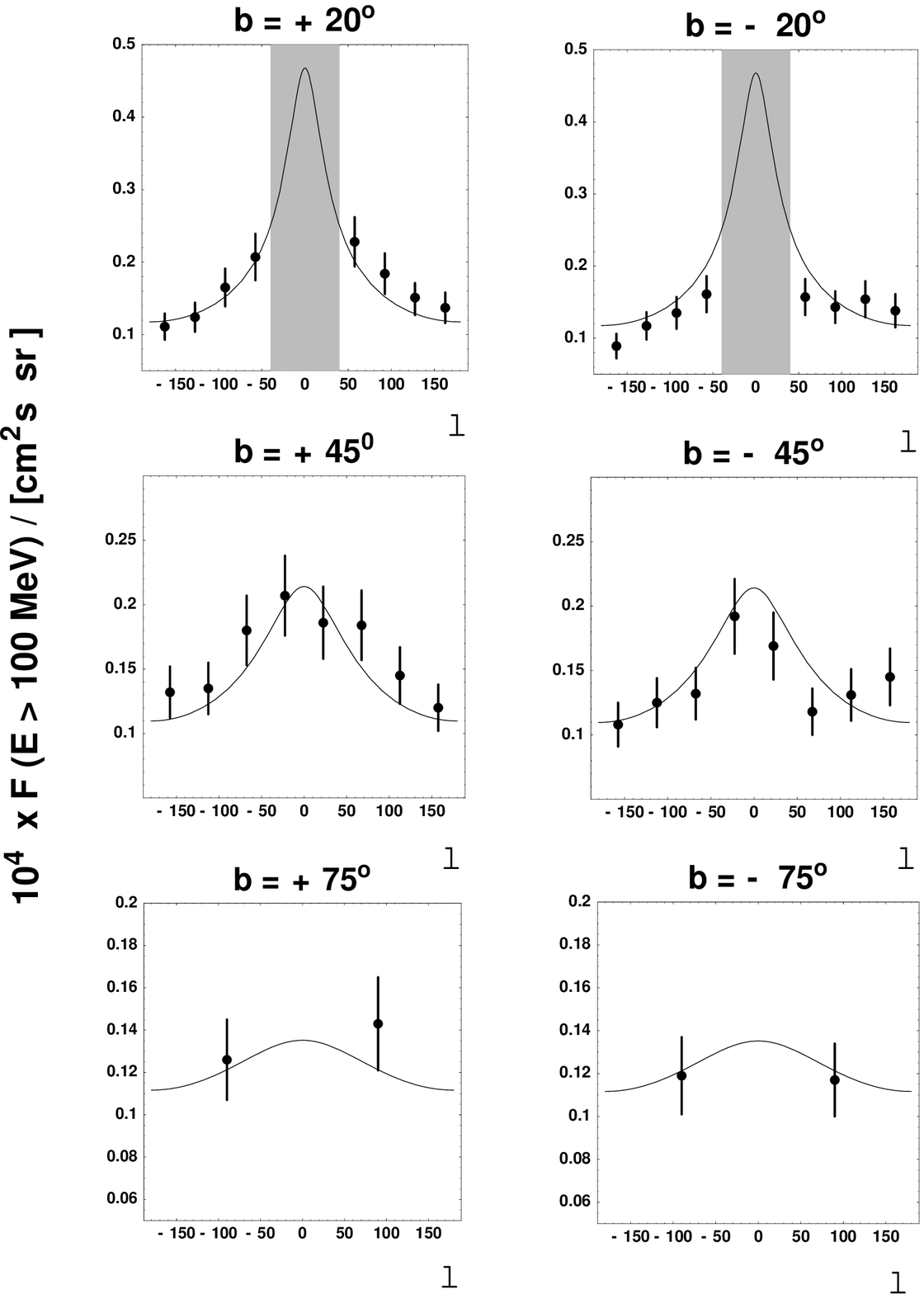,width=12cm}
%\vspace*{-14.6cm}
\caption{The flux of GBR photons above 100 MeV: comparison
between EGRET data and our model 
for ${\rm h_e=20}$ kpc, ${\rm \rho_e=35}$ kpc, as functions
of latitude at various fixed longitudes. The grey domain
is EGRET's mask.}
\vspace*{-0.5cm}
\end{center}
\end{figure}

\end{document}